\documentstyle[prl,aps,twocolumn,epsf]{revtex}
\def\break#1{\pagebreak \vspace*{#1}}
\begin{document}
\draft
\title{Crossover from coherent to incoherent dynamics in damped
quantum systems}
\author{
Reinhold Egger$^1$, Hermann Grabert$^1$, and Ulrich Weiss$^2$}
\address{
${}^1$Fakult\"at f\"ur Physik, Albert-Ludwigs-Universit\"at, 
Hermann-Herder-Strasse 3, D-79104 Freiburg, Germany\\
${}^2$Institut f\"ur Theoretische Physik, Universit\"at Stuttgart,
 Pfaffenwaldring 57, D-70550 Stuttgart, Germany}
\maketitle
\widetext
\begin{abstract}
The destruction of quantum coherence by environmental influences is
investigated 
taking the damped harmonic oscillator and the dissipative two-state system 
as prototypical examples.
It is shown that the location of the coherent-incoherent 
transition depends to a large degree on the dynamical
quantity under consideration.
\end{abstract}

\narrowtext

The destruction of quantum coherence by dissipative 
influences continues to be a problem of central interest in
atomic physics, condensed matter physics, and chemical physics.
It is relevant to phenomena as diverse as 
wave packet dynamics in atoms \cite{at},
defect tunneling in solids \cite{gw}, 
electron transfer in chemical and biological reactions \cite{chandler},
or quantum computers \cite{garg}.
Important insight into the effects of an environment
on quantum coherence was obtained in the last decade
based on studies of simple quantum systems in contact
with a heat bath \cite{leggett,weiss}.  
Very recently, focusing on the archetypical two-state system,
two groups \cite{costi,saleur} arrived at a result which, 
at first glance, seems to invalidate earlier studies.
They found that quantum coherence is destroyed at significantly
smaller damping strengths than thought previously, typically
differing by a factor $2/3$.  

In this Letter, we show that the coherent-incoherent transition 
depends on the particular dynamical quantity under 
consideration (e.g., correlation function, 
occupation probability, etc.). Since different dynamical 
quantities may be associated with different initial preparations 
of the system, quantum coherence may be more or less
sensitive to dissipation.   
The resulting critical value of the damping strength then changes to a
surprisingly large degree with the respective coherence criterion.
This will be explicitly demonstrated for the two fundamental dissipative 
quantum systems, namely the damped harmonic oscillator and
the dissipative two-state system. As the differences are most pronounced
at zero temperature, we confine ourselves to this limit.

We start with the exactly solvable case 
of a {\em harmonic oscillator}\, subject
to ohmic damping \cite{weiss}.
The respective classical equation of motion for the position $q(t)$ is
\begin{equation}\label{eqm}
\ddot{q}(t) + \gamma \dot{q}(t) + \omega_0^2 q(t) =  F(t)/M \; ,
\end{equation}
where $\gamma$ is the usual ohmic damping rate,
$\omega_0$  the frequency of the bare oscillator, and $M$ the mass
of the particle.  The response to the external
force $F(t)$ is described by 
\break{0.9in}
\begin{equation}\label{qexp}
\langle q(t) \rangle = \frac{1}{M\omega_0} \int_{-\infty}^t \, dt^\prime 
\, \chi_{\rm osc}(t-t^\prime)F(t^\prime)\;,
\end{equation}
where $\langle q(t) \rangle$ denotes the expectation value,
and $\chi_{\rm osc}(t)$ is 
the linear response function, which for the harmonic oscillator 
coincides with the classical response function.

One dynamical quantity of interest is the {\em equilibrium} 
correlation function 
\begin{equation}\label{cto}
C_{\rm osc}(t)= {\rm Re} \,\langle q(t) q(0) \rangle\;,
\end{equation}
where the brackets denote a ground-state average.
Consider now the associated spectral function \cite{costi}
\[
S_{\rm osc}(\omega) = C_{\rm osc}(\omega)/|\omega|\; ,
\]
which at $T=0$ is related to the absorptive part of the dynamical
susceptibility  by (we put $\hbar=1$) 
\[
S_{\rm osc}(\omega) = {\rm Im}\,\chi_{\rm osc}(\omega)/\omega \;.
\]
It is an even function of $\omega$, and the second form follows from
the fluctuation-dissipation theorem. From Eq.~(\ref{eqm}), 
the dynamical susceptibility 
\begin{equation} \label{cw}
\chi_{\rm osc}^{}(\omega) = 
\frac{\omega_0}{\omega_0^2-\omega^2-i\gamma\omega}\;,
\end{equation}
and hence the spectral function 
 \begin{equation}\label{st}
S_{\rm osc}(\omega) = \frac{\gamma\omega_0}{(\omega_0^2-\omega^2)^2 
+ \gamma^2\omega^2}  \;.
\end{equation}

Another problem of interest is the relaxation of the expectation value 
$\langle q(t) \rangle$ starting from a {\em nonequilibrium}
initial state.  Applying the force  
 $F(t)=M \omega_0^2 q_0\Theta(-t)$, the initial condition 
$\langle q(0)\rangle = q_0$ is prepared [cf.~Eq.~(\ref{qexp})],
 and the relevant dynamical quantity is
\begin{equation}\label{pto}
P_{\rm osc}(t)=\langle q(t)\rangle/q_0 \;.
\end{equation}
With the Fourier transform (\ref{cw}) of the response function,
one obtains from Eq.~(\ref{qexp})
\begin{equation}\label{qt}
 P_{\rm osc}(t)
= \cos(\Omega t - \phi) \exp(-\gamma t/2)/\cos(\phi) \; ,
\end{equation}
where $\Omega =\sqrt{\omega_0^2 - \gamma^2/4}$ and 
$\phi = \arctan(\gamma/2\Omega)$. 

Let us now discuss the coherent-incoherent
transition as a function of the dimensionless damping strength 
\[
\alpha = \gamma/2\omega_0\;.
\]
We first consider the appropriate coherence criterion based
on the spectral function $S_{\rm osc}(\omega)$.
For weak damping, $\alpha < \alpha_c$, the function $S_{\rm osc}(\omega)$ 
exhibits two inelastic 
peaks at finite frequency $\omega=\pm \omega_m(\alpha)$. 
At the critical damping strength $\alpha_c$, one
has $\omega_m(\alpha_c)=0$, and the two peaks merge into
a single quasi-elastic peak centered at $\omega=0$.
This quasi-elastic peak then persists for $\alpha>\alpha_c$. 
The value of $\alpha_c$ can be determined by inspecting the
sign of the curvature of $S_{\rm osc}(\omega)$ at zero 
frequency. From Eq.~(\ref{st}), we have the
small--$\omega$ expansion
\begin{equation}\label{osc}
 S_{\rm osc}(\omega) = 2 \alpha\chi_0^{2} \left[ 1 +
 (2 - 4\alpha^2)\chi_0^2\,\omega^2 +{\cal O}(\omega^4)
\right] \; ,
\end{equation}
where $\chi_0^{}=1/\omega_0$ is the static susceptibility.
Thus the curvature of $S_{\rm osc}(\omega)$
is positive (implying coherence) 
for $ \alpha < 1/\sqrt{2}$, but changes sign at the critical value.
Regarding the equilibrium correlations $C_{\rm osc}(t)$,
quantum coherence is suppressed for $\alpha>\alpha_c =1/\sqrt{2}$.

A different coherence criterion can be developed based
on the quantity $P_{\rm osc}(t)$.  
For weak damping, one finds damped oscillations such 
that $P_{\rm osc}(t)$ changes sign occasionally. 
As the damping strength $\alpha$ is increased,
$P_{\rm osc}(t)$ shows a transition from  damped
oscillatory to a purely incoherent behavior.
The coherent-incoherent transition is reached at the 
critical value $\alpha = \alpha_c^*$.
For $\alpha >\alpha_c^*$, one has $P_{\rm osc}(t) \geq 0$ 
for all times $t$. This coherence criterion based on a
nonequilibrium initial preparation has been
used in most previous work \cite{leggett,weiss}.
For the damped harmonic oscillator, we
see from Eq.~(\ref{qt}) that 
$P_{\rm osc}(t)$ becomes overdamped for 
$\alpha > 1$, while damped oscillations persist
for $\alpha<1$. Regarding the nonequilibrium quantity $P_{\rm osc}(t)$,
one has  destruction of quantum
coherence for $\alpha>\alpha_c^* =1$.
The ratio between the critical damping strengths
for the two dynamical quantities $C_{\rm osc}(t)$ and
$P_{\rm osc}(t)$ is then given as
\begin{equation}\label{hos}
\alpha_c/\alpha^*_c = 1/\sqrt{2} \; .
\end{equation}

We now turn to the case of a symmetric
{\em two-state system}. Coupling to 
an ohmic heat bath is described in terms of the spin-boson
Hamiltonian \cite{leggett,weiss}
\[
H= -(\Delta/2) \sigma_x + \sum_i \left( \frac{p_i^2}{2 m_i} + 
\frac{m_i \omega_i^2 x_i^2}{2} - \frac{c_i x_i \sigma_z}{2} \right) \;.
\]
The two eigenstates of $\sigma_z$ with eigenvalues $\pm 1$
are coupled by the transfer
matrix element $\Delta$ representing the tunnel splitting
of the free system.
The ohmic bath of harmonic oscillators
is fully characterized by the spectral density 
\[
 J(\omega) = \frac{\pi}{2}\sum_i \frac{c_i^2}{m_i \omega_i}
\delta(\omega-\omega_i) = 2 \pi \alpha  \omega \exp(-\omega/\omega_c) \;,
\]
where $\alpha$ is a dimensionless damping strength, 
and $\omega_c$ is a high-frequency cutoff.

An important aspect of this model is its correspondence
with the Kondo effect. 
By writing the partition function in the Coulomb gas representation,
a firm equivalence with the anisotropic Kondo model 
can be established \cite{anderson}.
This correspondence has been exploited  by 
Costi and Kieffer \cite{costi}, who applied a dynamical version of
Wilson's numerical renormalization group to the anisotropic Kondo model.
Furthermore, Lesage, Saleur, and Skorik\cite{saleur} 
have employed integrability and form-factor techniques
to study this model as well.
Both groups have calculated the ground-state spin-spin correlation
function of the Kondo model 
equivalent to the $T=0$ equilibrium two-state correlation function 
\[
C(t) = {\rm Re}\, \langle \sigma_z (t) \sigma_z (0) \rangle\;.
\]
This quantity is the direct analog of the harmonic oscillator
correlation function $C_{\rm osc}(t)$ defined in Eq.~(\ref{cto}).

Another useful quantity, particularly in the context of
macroscopic quantum coherence \cite{garg,leggett}, 
is the occupation probability
\[
P(t) = \langle \sigma_z (t) \rangle \;,
\]
corresponding to the harmonic oscillator quantity
$P_{\rm osc}(t)$ defined in Eq.~(\ref{pto}).
In contrast to $C(t)$, the function $P(t)$ is subject to the nonequilibrium
initial preparation $\sigma_z(t=0)=+1$.
This preparation of the initial state may be realized by applying
a large external bias.  Thereby the spin is held fixed in the state 
$\sigma_z=+1$ with equilibrated environment.
At time zero, the constraint is released, and the dynamics
starts out from $P(0)=1$ with 
this factorized system-environment initial state.

Let us now discuss the coherent-incoherent transition for
the dissipative two-state system, starting with the 
equilibrium correlations. Similar to Eq.~(\ref{osc}),
the function $S(\omega)=C(\omega)/|\omega|$
has the low-frequency expansion
\begin{equation}\label{expan1}
S(\omega) = 2\pi \alpha \chi_0^2 [ 1 + \kappa(\alpha) \chi_0^2\,\omega^2 + 
{\cal O}(\omega^4)] \; ,
\end{equation} 
where $\chi_0^{}$ is the static susceptibility
and $\kappa(\alpha)$ is a dimensionless parameter.
The zero-frequency limit of Eq.~(\ref{expan1})
is the generalized Shiba relation for the spin-boson problem \cite{sass90}.  
As a result of this relation, 
spin-spin correlations decay asymptotically as $C(t)
=  - 2 \alpha \chi_0^2 /t^2$. In Refs.\cite{costi,saleur},
the function $\kappa(\alpha)$ has been computed and was found to change  
sign at the critical value $\alpha_c = 1/3$.

Next we analyze the transition from oscillations to
incoherent relaxation in the quantity $P(t)$.
Within the widely-used noninteracting-blip approximation (NIBA),
the critical value is $\alpha^*_c=1/2$ \cite{leggett}. 
This result can easily be seen
by switching to the Laplace transform $P(\lambda)$ and defining a
self-energy $\Sigma(\lambda)$, 
\begin{equation} \label{plambda}
P(\lambda)= 1/[\lambda+\Sigma(\lambda)] \;.
\end{equation}
NIBA gives for the self-energy \cite{leggett,weiss}
\begin{eqnarray}  \label{niba}
\Sigma (\lambda) &=&  \Delta^2 \cos(\pi \alpha) \int_0^\infty d\tau 
\, \frac{\exp(-\lambda \tau)}{(\omega_c \tau)^{2\alpha}} \\
&=& \Delta_e (\Delta_e/ \lambda)^{1-2\alpha} \;,\nonumber
\end{eqnarray}
with the effective frequency scale
\[
\Delta_e = \Delta  [ \cos (\pi \alpha)\Gamma(1-2\alpha)]^{1/2(1-\alpha)} 
(\Delta/\omega_c)^{\alpha/(1-\alpha)}  \;.
\]
In the limit $\alpha\to1/2$, the frequency $\Delta_e$ 
approaches $\pi\Delta^2/2\omega_c$.
For $\alpha>0$, the function $P(\lambda)$ has a branch point at $\lambda=0$.
The complex $\lambda$-plane is cut along the negative real axis, and in the
cut plane $P(\lambda)$ is single-valued. The cut leads to an incoherent
contribution to $P(t)$. Moreover, for $\alpha <1/2$, the integrand has a
conjugate pair of poles in the cut $\lambda$-plane describing damped
oscillations.

To study the coherent-incoherent transition, we 
consider the pole condition from Eq.~(\ref{plambda}),
\begin{equation}\label{poleniba}
(\lambda/\Delta_e)^{2-2\alpha} = -1 \;,
\end{equation}
and put 
\begin{equation}\label{eps}
\alpha=1/2-\epsilon\;,\quad |\epsilon|\ll 1\;.
\end{equation}
For $\epsilon > 0$ ($\alpha < 1/2$), insertion of the ansatz 
\[
\lambda/\Delta_e=-1+\epsilon u \pm i \epsilon v + {\cal O}(\epsilon^2) 
\]
into Eq.~(\ref{poleniba}) yields
$u=0$ and $v= 2\pi$. Hence, in the NIBA the damping rate $\Gamma$ and the
oscillation frequency $\Omega$ are up to corrections
of order $\epsilon^2$ 
\begin{equation} \label{nibac} 
\Gamma/ \Delta_e  = 1 \;, \qquad \Omega/\Delta_e=2\pi \epsilon\;.
\end{equation}
This results in damped oscillations with
frequency $\Omega$. In contrast, for $\epsilon<0$ ($\alpha>1/2$),
the poles are not in the cut plane and therefore give no contribution
to $P(t)$. In that case, $P(t)$ is completely given by the incoherent
branch-cut contribution. Thus the critical damping strength is indeed
$\alpha_c^*=1/2$.

Remarkably, exactly at the special value $\alpha=1/2$, NIBA becomes 
exact\cite{leggett}, while it is only an approximation 
for $\alpha \neq 1/2$. One serious deficiency comes from the
branch-cut contribution which would imply
the existence of an algebraic long-time tail. From
Eq.~(\ref{niba}), for $\alpha=1/2-\epsilon$ and
$\lambda \to 0$, one finds $P(\lambda)\sim \lambda^{2\epsilon}$.
The asymptotic branch-cut contribution is therefore 
\begin{equation} \label{tail}
P(t) = -2\epsilon/(\Delta_e t)^{1+2\epsilon}  \;.
\end{equation}
This term decaying slower than $1/t^2$ contradicts the
fluctuation-dissipation theorem. Hence, the  long-time tail (\ref{tail})
is an unphysical artefact of NIBA\cite{leggett}.  Such a failure raises
the question whether the NIBA value $\alpha^*_c=1/2$
for the coherent-incoherent transition remains correct.

To investigate the coherent-incoherent transition
beyond NIBA, we now systematically expand around the
exactly  solvable case $\alpha=1/2$.
In diagrammatic terms, the exact
self-energy $\Sigma(\lambda)$ is the sum over all
irreducible arrangements of ``blips'' and ``sojourns''.
A blip (sojourn) refers
to the time spent in an off-diagonal (diagonal) state
of the reduced density matrix\cite{leggett,weiss}.
For instance, the NIBA expression (\ref{niba}) is just given by the
single-blip contribution, thereby
effectively disregarding all inter-blip interactions in $P(\lambda)$.
The fact that NIBA becomes exact
for $\alpha=1/2$ is explained simply by the concept of {\em collapsed blips}
\cite{weiss}. In view of Eq.~(\ref{eps}), there is a factor
$\cos(\pi\alpha)=\pi\epsilon$ in Eq.~(\ref{niba}).
This ${\cal O}(\epsilon)$ factor must be cancelled by a $1/\epsilon$
short-time contribution  of the $\tau$-integral over the
length of the blip such that a finite result can arise.
Therefore, for $\alpha \to 1/2$, only blips of effectively 
vanishing length, that is ``collapsed'' blips, contribute.
Since interactions among different collapsed blips vanish,  
NIBA becomes exact \cite{weiss}. 

For nonzero $\epsilon$, the blip length is finite, and 
one has to take into account all sequences of {\em collapsed sojourns} 
within an extended blip of length $\tau$. Their
presence crucially modifies the self-energy in the limit $\lambda\to 0$.
Since collapsed sojourns do not interact with blips or other 
sojourns, a grand-canonical gas of collapsed sojourns merely
gives a factor $\exp(-\Delta_e \tau/2)$, and 
the self-energy (\ref{niba}) is changed into  
\begin{eqnarray} \nonumber
\Sigma (\lambda) &=&  \Delta^2 \pi \epsilon \int_0^\infty d\tau \,
\frac{\exp[-(\lambda+\Delta_e/2)\tau]}{(\omega_c \tau)^{1-2\epsilon}} \\
&=& \Delta_e (\lambda/\Delta_e + 1/2)^{-2\epsilon} \;.\label{full}
\end{eqnarray}

With the regularized self-energy (\ref{full}), the pole condition 
now reads
\[
\lambda \left (1/2 + \lambda/\Delta_e \right)^{2\epsilon} = - \Delta_e\;.
\]
This yields the  exact decay rate and oscillation frequency up to 
corrections of order $\epsilon^2$,
\[
\Gamma/\Delta_e = 1-2\epsilon \ln 2  \;,\quad
\Omega/\Delta_e =  2\pi \epsilon\;.
\]
Compared to the NIBA result (\ref{nibac}), 
the rate acquires a correction in order $\epsilon$.
However, the oscillation frequency remains completely unchanged.  
Therefore, the NIBA value $\alpha_c^* = 1/2$ for the location 
of the coherent-incoherent transition turns out to be the 
{\em exact}\, result.  This implies the critical ratio
\[
 \alpha_c/\alpha^*_c = 2/3 \;,
\]
which is slightly smaller than the 
respective ratio (\ref{hos}) for the damped harmonic oscillator. 

The exact self-energy (\ref{full}) shifts the branch point of
$P(\lambda)$ and thereby removes the spurious algebraic
long-time tails (\ref{tail}). For $\epsilon >0$, the leading branch-cut 
contribution at long times  is now given by
\[
P(t) = - 2\epsilon \frac{\exp(-\Delta_e t/2)}{(\Delta_e t)^{1+2\epsilon}}  \;,
\]
while for $\epsilon <0$, we obtain
\[
P(t) = 8|\epsilon|
 \frac{\exp(-\Delta_e t/2)}{(\Delta_e t)^{1+2|\epsilon|}}  \;.
\]
Thus, the unphysical algebraic long-time tails are suppressed
by an exponential decay factor.  It is straightforward to see
that for $\epsilon>0$ ($\alpha<1/2$),
the full cut contribution is negative, while it becomes positive
for $\epsilon<0$ ($\alpha>1/2$). Indeed, for $\alpha\geq 1/2$, the
function $P(t)$ is positive and monotonically decaying, i.e.,
the dynamics is fully incoherent. In marked contrast to 
the NIBA result (\ref{tail}), the power of the algebraic decay factor
does not depend on the sign of $\epsilon$.

\begin{figure}
\epsfysize=6.5cm
\epsffile{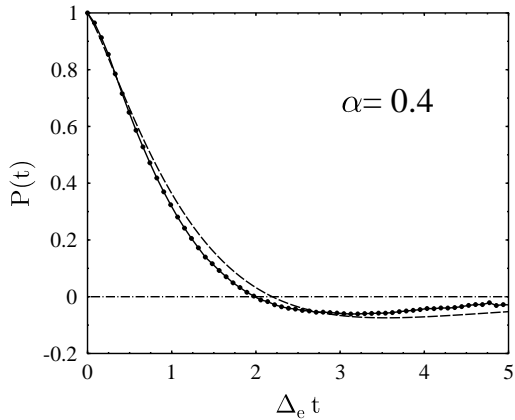}
\caption[]{\label{fig1} Monte Carlo data for $P(t)$ at $\alpha=0.4$ and $T=0$.
Circles are data points for $\Delta/\omega_c=1/6$,
the solid curve connecting them is a guide for the eye only.
Statistical errors are well below $5\%$. The dashed curve is the NIBA result. }
\end{figure}

The remarkable success of NIBA in predicting the
correct value of $\alpha_c^*$ provokes questions about 
the general quality of such a simple approximation
for the full range of $\alpha$.  A convenient tool
to investigate this issue is the real-time quantum Monte Carlo 
simulation method  \cite{egger}.
This technique permits a numerically exact calculation of
both dynamical quantities $C(t)$ and $P(t)$ by stochastic
evaluation of the respective 
real-time path-integral representations.
The dynamical sign problem arising from the interference between different
real-time paths can be largely circumvented by a partial 
summation scheme, and stable simulations can be carried out for rather
long times. The simulation code permits a computation of $P(t)$
directly at zero temperature, where the case $\alpha=1/2$ serves
as a convenient benchmark which is indeed passed accurately \cite{egger}.

Numerical results for $P(t)$ at $\alpha=0.4$
are shown in  Fig.~\ref{fig1}, and data for $\alpha\geq 1/2$ 
can be found in Ref.\cite{egger}.
These data represent universal scaling curves
in the sense that they can be obtained from different $\omega_c$ and/or
$\Delta$ by employing the effective frequency scale $\Delta_e$.
Apparently, on short-to-intermediate time scales,
NIBA yields a very accurate prediction for $P(t)$.
However, the prediction $C(t)=P(t)$ as well as the
long-time tails (\ref{tail}) are clear deficiencies 
of this approximation\cite{egger}.

In conclusion, we have demonstrated that the initial preparation
of a dissipative quantum system leads to drastic changes
regarding the transition from coherence to incoherence as the damping strength
is increased. Both the damped harmonic oscillator and the 
dissipative two-state system 
show that the two coherence criteria previously employed in the
literature lead to a factor $1/\sqrt{2}$ and $2/3$
difference in the critical damping strengths, respectively. Any
investigation of the environmental destruction of quantum coherence 
thus necessitates a clear specification of the physical quantities
under study.

RE and HG acknowledge support by the DFG-SPP
``Zeitabh\"angige Ph\"anomene und Methoden in Quan\-ten\-systemen
der Physik und Chemie''. UW is supported by the DFG-SFB 382 ``Verfahren
und Algorithmen zur Simulation physikalischer Prozesse auf
H\"ochstleistungsrechnern''.  Finally, 
RE and UW wish to acknowledge the hospitality of the Institute for
Scientific Interchange (ISI) Foundation, Torino, where 
parts of this work were carried out under contract
ERBCHRX-CT920020.

\end{document}